\documentclass[runningheads]{llncs}
\usepackage[T1]{fontenc}
\usepackage{graphicx}
\usepackage{subcaption}
\usepackage{textcomp}

\begin{document}
\title{Top-Down vs. Bottom-Up Approaches for Automatic Educational Knowledge Graph Construction in CourseMapper}
\titlerunning{Top-Down vs. Bottom-Up EduKG Construction in CourseMapper}
%
\author{
Qurat Ul Ain \and
Mohamed Amine Chatti \and
Amr Shakhshir \and
Jean Qussa \and
Rawaa Alatrash \and
Shoeb Joarder
}
\authorrunning{Ain et al.}

\institute{
Social Computing Group, Faculty of Computer Science, University of Duisburg-Essen, Germany
}
\maketitle              
\begin{abstract}
The automatic construction of Educational Knowledge Graphs (EduKGs) is crucial for modeling domain knowledge in digital learning environments, particularly in Massive Open Online Courses (MOOCs). However, identifying the most effective approach for constructing accurate EduKGs remains a challenge. This study compares Top-down and Bottom-up approaches for automatic EduKG construction, evaluating their effectiveness in capturing and structuring knowledge concepts from learning materials in our MOOC platform CourseMapper. Through a user study and expert validation using Simple Random Sampling (SRS), results indicate that the Bottom-up approach outperforms the Top-down approach in accurately identifying and mapping key knowledge concepts. To further enhance EduKG accuracy, we integrate a Human-in-the-Loop approach, allowing course moderators to review and refine the EduKG before publication. This structured comparison provides a scalable framework for improving knowledge representation in MOOCs, ultimately supporting more personalized and adaptive learning experiences.
\end{abstract}
\section{Introduction}
The rapid growth of online education has led to the widespread adoption of Massive Open Online Courses (MOOCs), offering learners open access to high-quality education at scale and fostering lifelong learning opportunities \cite{10.1145/3579991}\cite{ABUSALIH2024e25383}. As MOOCs continue to evolve, Artificial Intelligence (AI) is playing a transformative role in enhancing their effectiveness. Among the various AI-driven innovations in education, Knowledge Graphs (KGs) have emerged as a powerful tool for structuring and organizing knowledge, and enabling personalized and interconnected learning experiences. Their application in education, referred to as Educational Knowledge Graphs (EduKGs), is revolutionizing how knowledge is organized, represented, and applied, ultimately enriching the learning experiences \cite{ABUSALIH2024e25383}.

EduKGs are increasingly being integrated into MOOCs for various purposes, e.g. optimizing learning resource utilization \cite{dang2019mooc}, predicting learning behavior \cite{Xia14012025}, recommending knowledge concepts and courses \cite{10.1145/3579991}\cite{10.1117/12.2682468}, and many more. Despite their benefits, constructing accurate EduKGs remains a significant challenge. Traditional methods depend on domain experts, making EduKGs construction time-consuming and resource-intensive \cite{info14100526}. Moreover, the increasing volume of educational data has driven the need for automated EduKG construction, yet existing approaches often struggle with accuracy and performance \cite{Manrique2018knowledge}\cite{info14100526}. Recent advancements in Large Language Models (LLMs) have driven research into enhancing EduKG generation with LLM-based approaches as well \cite{jhajj2024educational}. However, there is currently no standard approach for constructing EduKGs in MOOCs, particularly in terms of evaluating different methodologies such as Top-down and Bottom-up. Identifying the most effective strategy is crucial, as the accuracy of EduKGs directly impacts their usefulness in MOOCs. Moreover, given that MOOCs consist of multiple materials structured into pages, it is essential to explore whether EduKGs should be constructed holistically from entire materials or incrementally page-by-page. To address this gap, in this paper, we experiment with two pipelines of automatic EduKG construction, namely Top-down and Bottom-up approaches in our MOOC platform CourseMapper \cite{ain2022learning}. Our findings indicate that the Bottom-up approach achieves the highest accuracy, demonstrating its effectiveness in constructing reliable EduKGs for MOOCs. Additionally, acknowledging the importance of human involvement in EduKG construction, we integrate a Human-in-the-Loop approach in our pipeline, enabling course experts to review and refine the automatically generated EduKGs before publication. This balances automation with human expertise, ensuring quality while minimizing manual effort.
\begin{figure}[h]
    \centering
    \begin{subfigure}{0.49\textwidth}  
        \includegraphics[width=\linewidth]{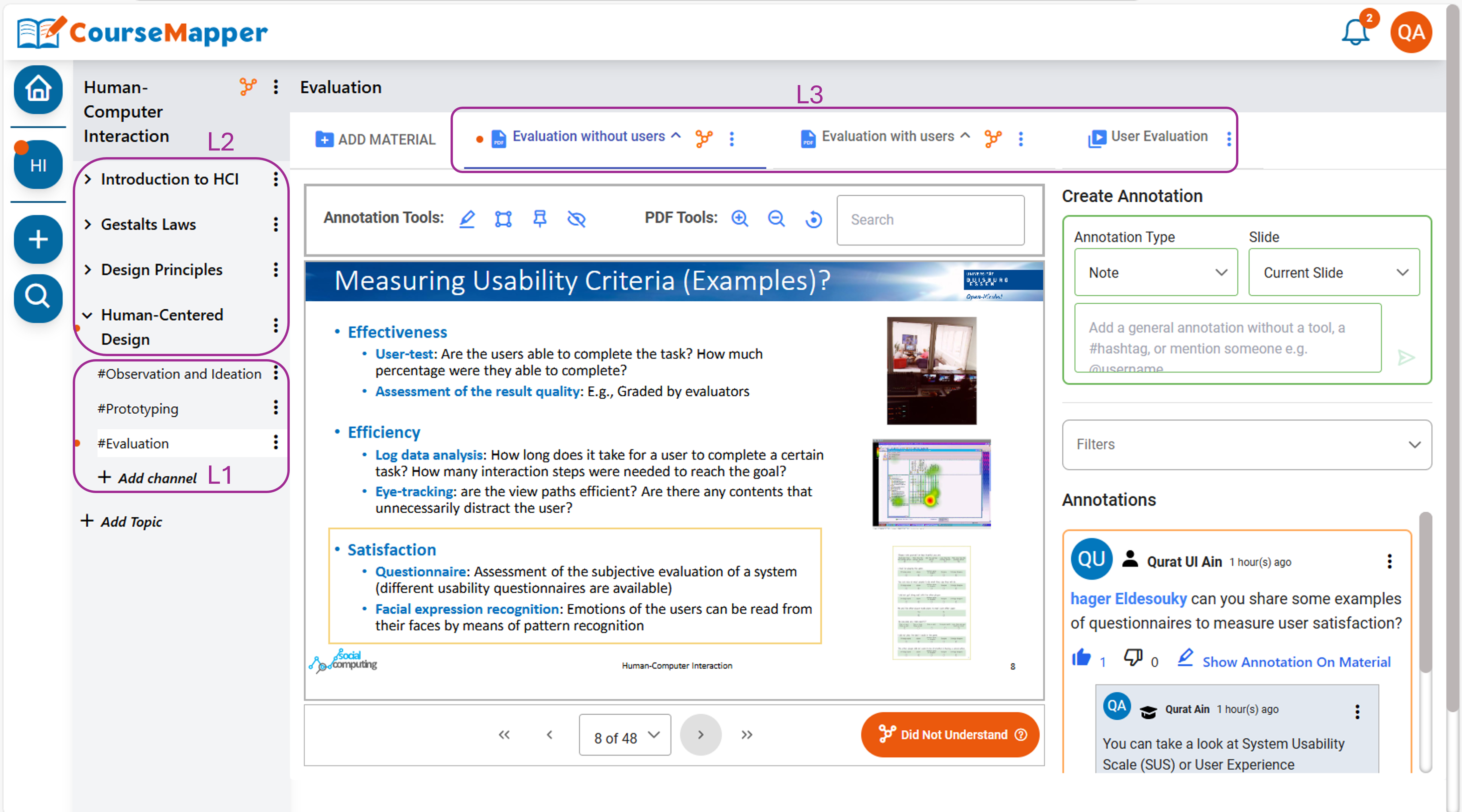}
        \caption{Learning Channels}
        \label{lc}
    \end{subfigure}
    \hfill
    \begin{subfigure}{0.49\textwidth}
        \includegraphics[width=\linewidth]{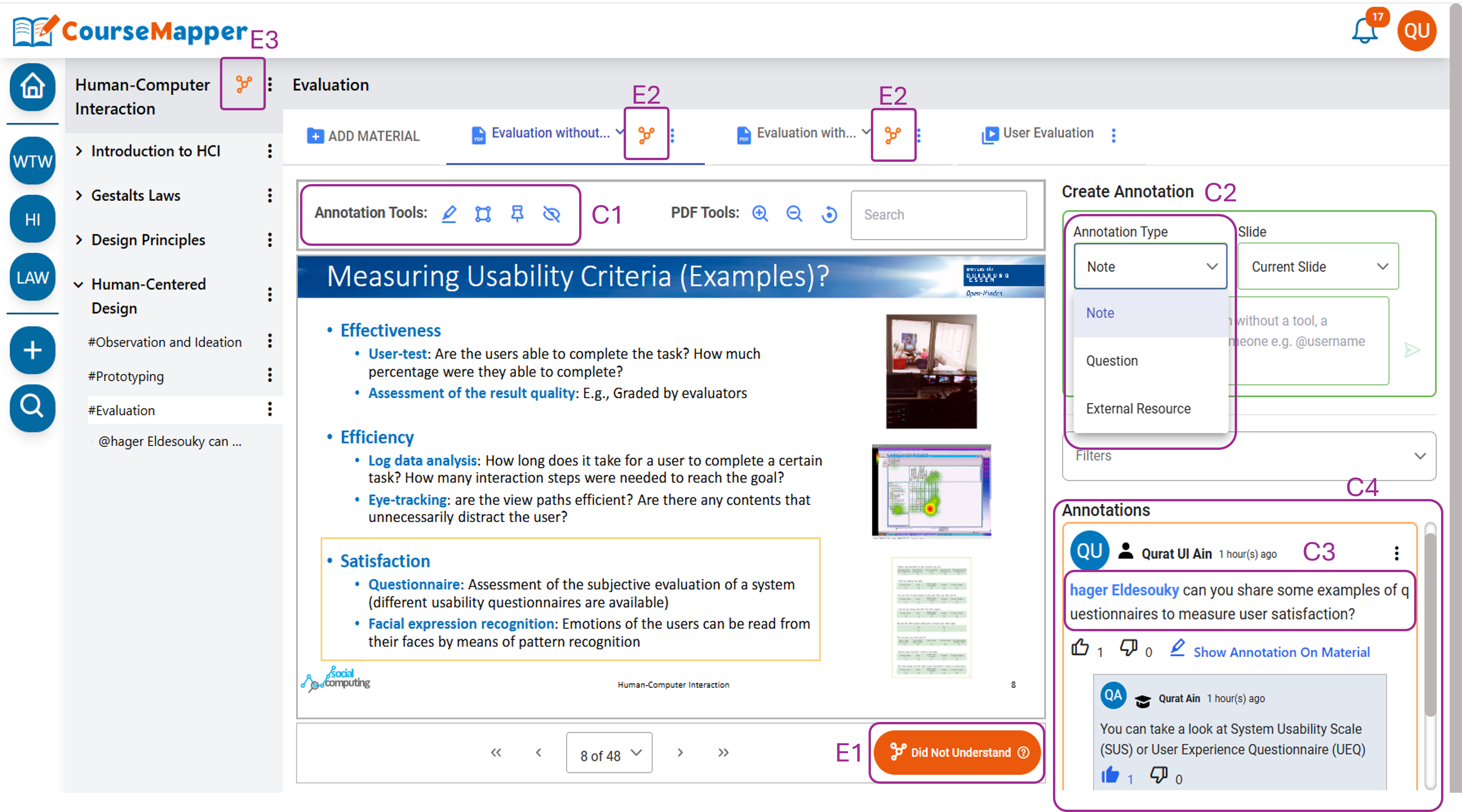}
        \caption{Collaboration and Communication}
        \label{cc}
    \end{subfigure}
    \begin{subfigure}{0.49\textwidth}
        \includegraphics[width=\linewidth]{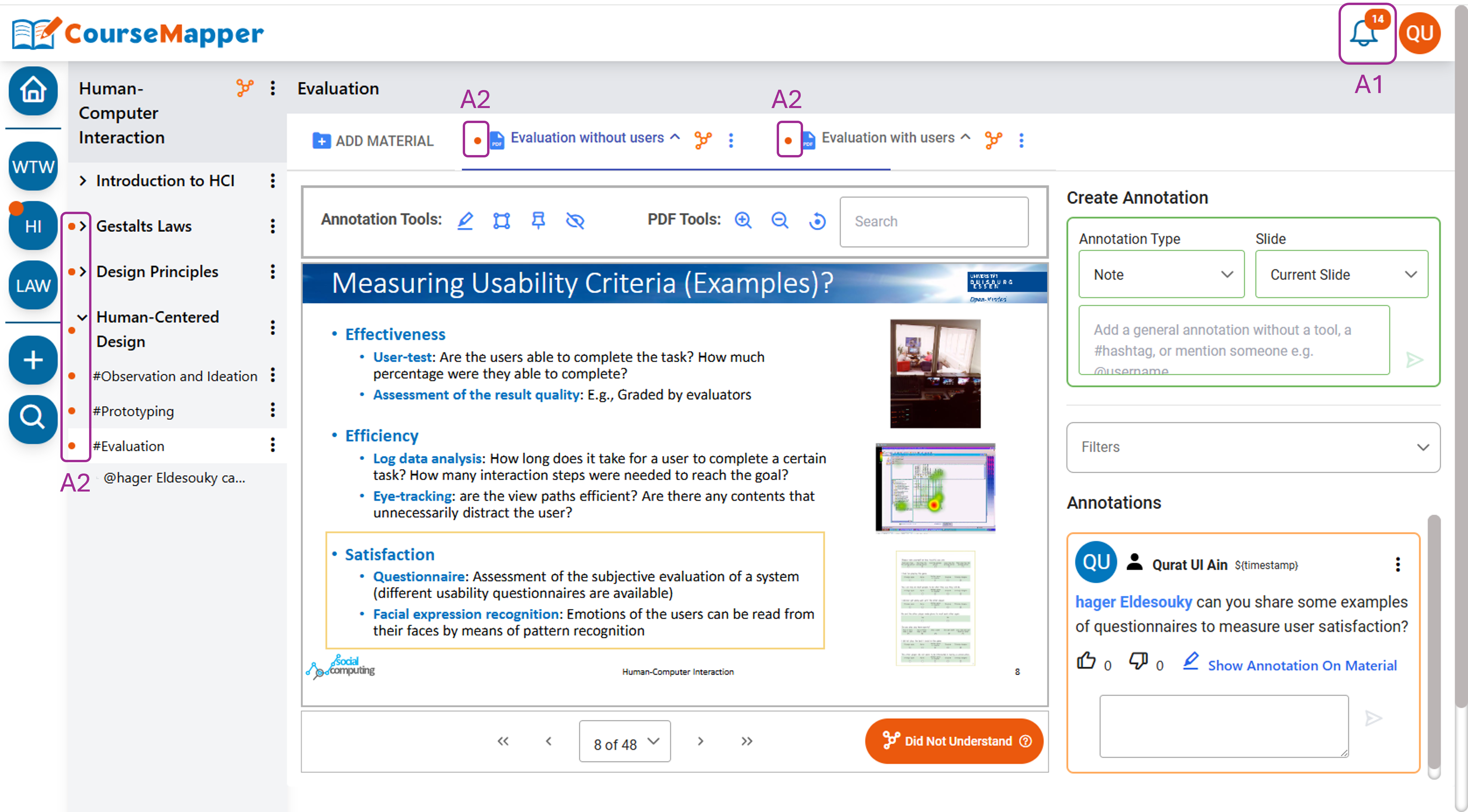}
        \caption{Awareness}
        \label{aw}
    \end{subfigure}
    \hfill
    \begin{subfigure}{0.49\textwidth}
        \includegraphics[width=\linewidth]{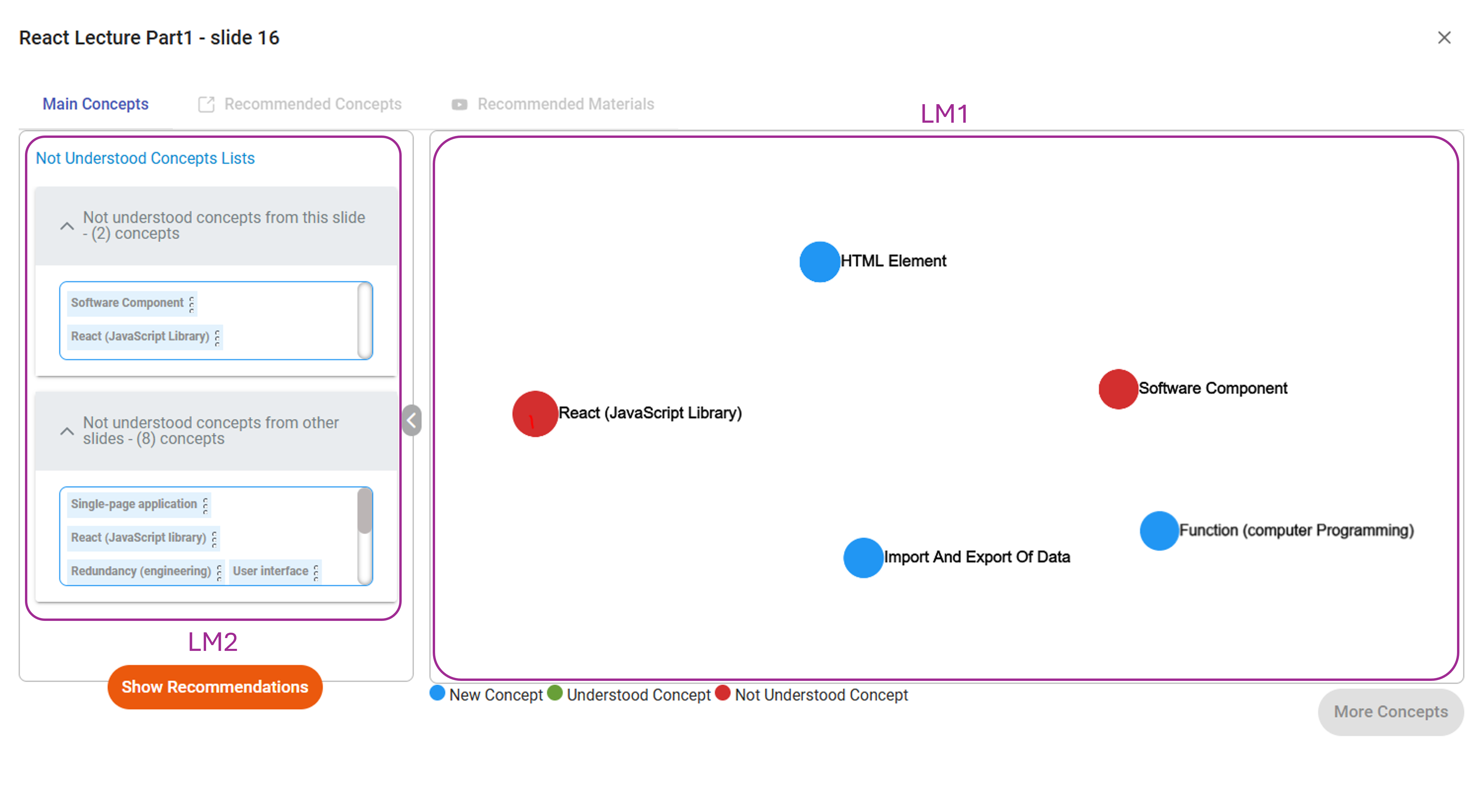}
        \caption{Learner Modeling}
        \label{lm}
    \end{subfigure}
    \begin{subfigure}{0.49\textwidth}
        \includegraphics[width=\linewidth]{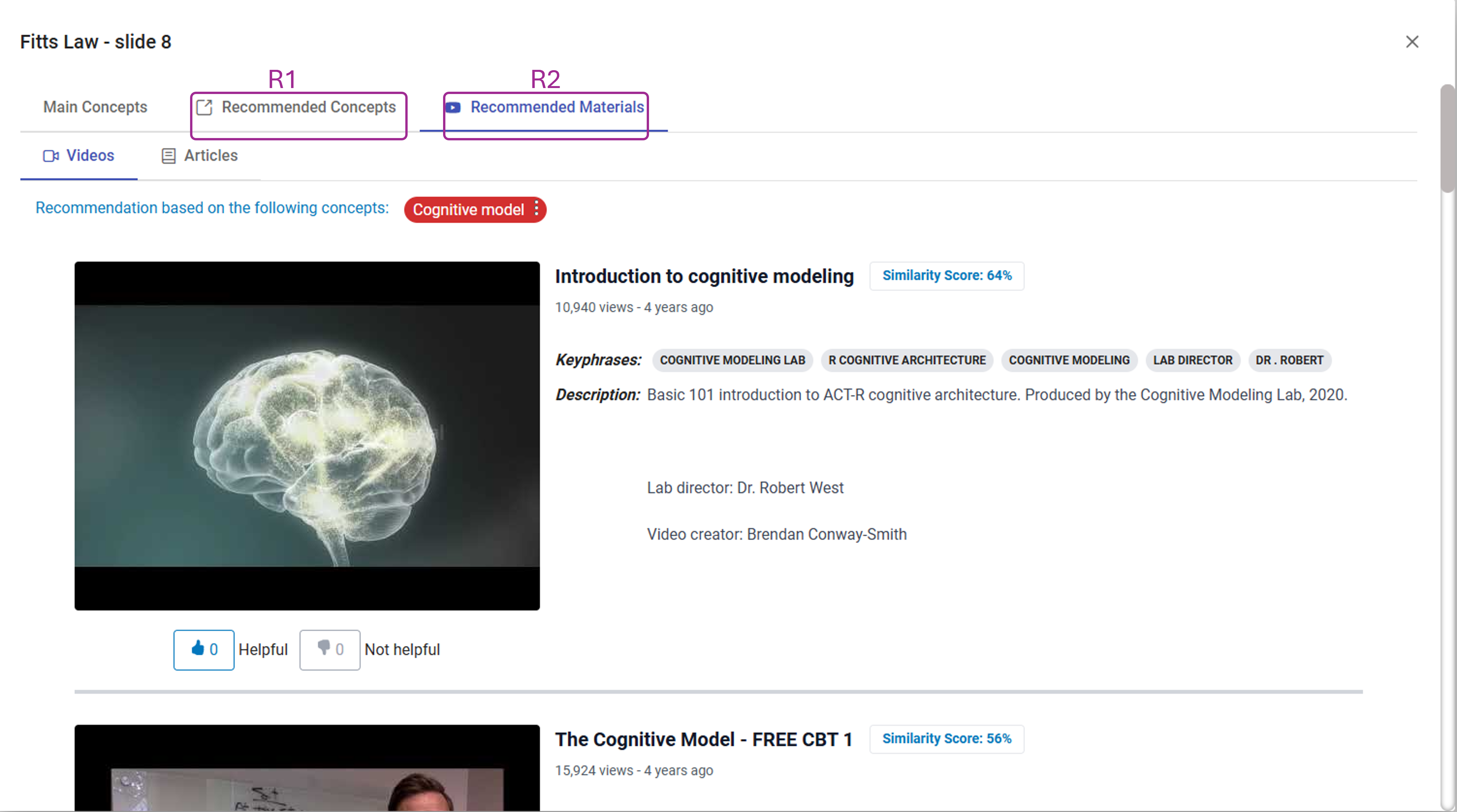}
        \caption{Recommendation}
        \label{r}
    \end{subfigure}
    \hfill
    \begin{subfigure}{0.49\textwidth}
        \includegraphics[width=\linewidth]{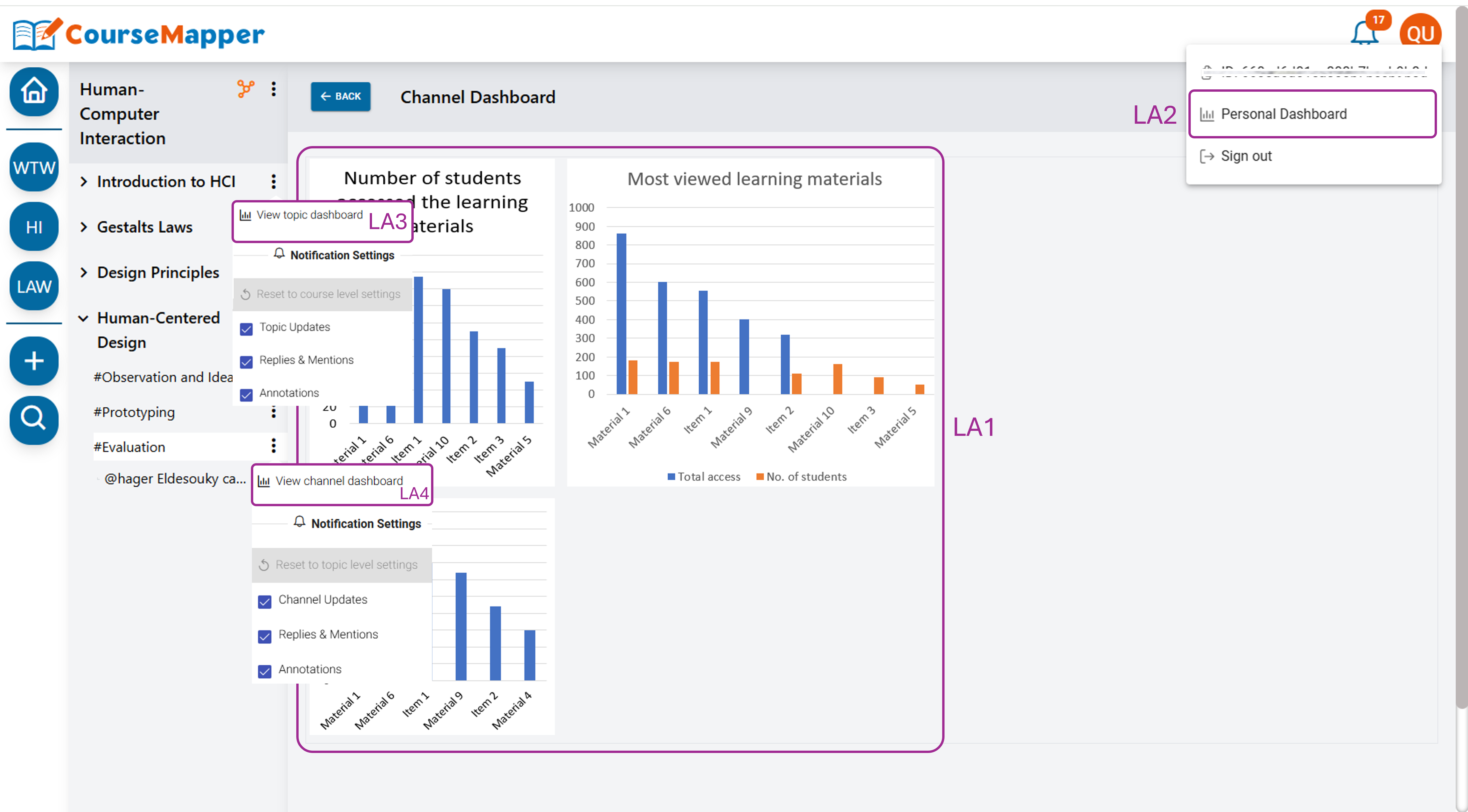}
        \caption{Learning Analytics}
        \label{la}
    \end{subfigure}
    \caption{An overview of UI of CourseMapper demonstrating different features}
    \label{fig:screenshots}
\end{figure}
\section{CourseMapper} \label{blinded}
Our MOOC platform CourseMapper consists of a range of unique features designed to enhance the online learning experience by addressing various learner needs. These features set our platform apart from existing MOOC platforms by offering innovative functionalities that improve interaction and communication in online learning, foster learner engagement, enhance personalization, and support learning analytics (see Figure \ref{fig:screenshots}).
\paragraph{\textbf{Learning Channels:}}
Each course in CourseMapper includes multiple learning channels, which serve as collaborative spaces within the MOOC platform. These learning channels (Figure \ref{lc}, L1) are created for each course topic (Figure \ref{lc}, L2), enabling learners to engage with PDF and video learning materials (Figure \ref{lc}, L3), discuss concepts with peers and instructors, and share relevant resources within the designated space. The concept of learning channels provides a more organized and interactive way to structure courses, fostering deeper engagement and knowledge sharing among learners.
\paragraph{\textbf{Collaboration and Communication:}}
Learners can collaborate on PDF and video learning materials using three different annotation tools (i.e., highlight, draw, pinpoint) (Figure \ref{cc}, C1) to mark specific parts of a learning material and add a note, question, or external resource link, referred to as annotation types (Figure \ref{cc}, C2). Additionally, the mention feature (using @) (Figure \ref{cc}, C3), allows learners to tag others by name, enabling direct interaction. All annotations appear in the discussion panel (Figure \ref{cc}, C4) alongside the learning material, where learners can view, respond to, or like/dislike them. Using shared annotations, learners can engage in deeper discussions on learning materials, enhance collaboration, and improve communication with both peers and instructors.
\paragraph{\textbf{Awareness:}}
CourseMapper includes a notification system to enhance awareness of course activities among course participants. Learner's activities (e.g., annotations, replies etc.) are logged as xAPI statements and then used to generate relevant notifications. Based on their personalized settings, learners receive tailored notifications in their newsfeed (Figure \ref{aw}, A1), categorized into course updates, replies and mentions, and annotations.
In addition to notifications, an orange dot indicator (Figure \ref{aw}, A2) serves as a visual cue, highlighting the respective course, topic, learning channel, and/or learning material whenever a new activity occurs. This feature helps learners stay informed and engaged by drawing their attention to recent course updates.
\paragraph{\textbf{Educational Knowledge Graphs:}}
In CourseMapper, Educational Knowledge Graphs (EduKGs) provide learners a structured overview of key concepts and their relationships. EduKGs are built at three levels: Slide-EduKG (concepts within a slide) (Figure \ref{cc}, E1), LM-EduKG (concepts across a learning material) (Figure \ref{cc}, E2), and Course-EduKG (concepts throughout a course) (Figure \ref{cc}, E3). Stored in a Neo4j graph database, nodes represent concepts, categories, slides, and learning materials, connected by edges representing the relationships between them.
EduKGs construction details are disucssed in Section \ref{oldpipe}.
\paragraph{\textbf{Learner Modeling:}}
Learner modeling plays a crucial role in enhancing personalization, engagement, and learning effectiveness in MOOCs. In CourseMapper, each PDF learning material includes a "Did Not Understand (DNU)" button at the bottom (Figure \ref{cc}, E1). When clicked, learners are presented with the Slide-EduKG containing the top five main concepts extracted from the content of the current page (Figure \ref{lm}, LM1). They can then mark the concepts they do not understand (DNU) (Figure \ref{lm}, LM2), allowing them to explicitly communicate their knowledge state to the system rather than having the system infer it implicitly based on their behavioral data. These DNU concepts are linked to the learner to formulate their Personal Knowledge Graph (PKG), creating a structured representation of the learner. This PKG-based learner model is further leveraged to provide personalized recommendations.
\paragraph{\textbf{Recommendation:}}
PKG-based learner models are further used to generate personalized recommendations of related knowledge concepts (Figure \ref{r}, R1), using Graph Convolutional Networks (GCNs) and pre-trained transformer language model encoders. To increase transparency, explanations of the recommended concepts are provided using the structural and semantic information in the EduKG \cite{ALATRASH2024100193}. Moreover, the learners are provided personalized recommendations of external learning resources (Figure \ref{r}, R2) including YouTube videos and Wikipedia articles \cite{QURATlak}, using both PKG-based and content-based recommendation algorithms.
\paragraph{\textbf{Learning Analytics:}}
In CourseMapper, learners' activity data collected as xAPI statements, is used to analyze user engagement patterns and generate meaningful learning insights through an external open Learning Analytics (LA) platform, OpenLAP \cite{Joarder2024NoCode}. OpenLAP supports self-service LA by empowering end-users to take control of the LA indicator design process, through intuitive user interfaces. Using OpenLAP, various LA indicators are generated from the xAPI data (Figure \ref{la}, LA1) and visually represented in dashboards at different levels (Figure \ref{la}, LA2-3-4) within CourseMapper through iframes. These analytics enable learners and educators to track progress, identify learning patterns, and make informed decisions to improve the learning experience.
\section{EduKG Construction Phases} \label{pipelineoverview}
EduKG construction is a multi-phase process involving several key steps described below.
\paragraph{\textbf{Text Extraction:}}\label{textext}
 This phase extracts text from PDF learning materials while preserving document structure. Standard extraction methods often overlook layout variations, so we use a simplified layout-aware approach. It involves: (1) identifying contiguous text blocks, (2) categorizing them using rules, and (3) merging them in sequence. We used PDFMiner \cite{unixuserPDFMiner} that retrieves character positions, grouping them into structured text blocks based on coordinate proximity.
\paragraph{\textbf{Keyphrase Extraction:}}
After extracting text, we apply keyphrase extraction as a pre-step for entity linking to Wikipedia. This approach improves efficiency by reducing the volume of text sent to the entity-linking service. 
For keyphrase extraction algorithm, we used \begin{math}SIFRank_{SqueezeBERT}\end{math} \cite{info14100526} chosen based on its accuracy and performance results \cite{info14100526}. 
\paragraph{\textbf{Concept Identification:}} \label{CI}
Extracted keyphrases are mapped to relevant concepts from external knowledge base DBpedia. Following \cite{Manrique2018knowledge}, we use DBpedia Spotlight \cite{mendes2011dbpedia} to link keyphrases to DBpedia concepts through spotting, candidate selection, and disambiguation, with the support value set to 5 and the confidence threshold to 0.35 \cite{grevisse2018knowledge}. However, entity-linking tools like DBpedia Spotlight can produce incorrect annotations due to automatic processing without manual verification. To mitigate this, we apply a weighting strategy to assess semantic similarity between identified concepts and learning materials, described later.
\paragraph{\textbf{Concept Expansion:}} \label{expansionold}
To improve EduKG coverage, diversity, and knowledge exploration, we expand identified concepts, as keyphrase extraction and concept identification may miss some relevant concepts \cite{manrique2018weighting}. This expansion follows two approaches: related concept expansion, which enriches EduKG with semantically related DBpedia concepts (e.g., linking “Natural language processing” to “Natural language understanding” via dbo:wikiPageWikiLink), and category-based expansion, which associates concepts with their DBpedia categories (e.g., linking “Natural language processing” to “Category:Computational linguistics” via dct:subject) using SPARQL queries, providing hierarchical context and facilitate broader concept discovery. 
\paragraph{\textbf{Concept Weighting:}} \label{weighting}
While concept expansion enriches the EduKG, it may introduce noise by adding irrelevant concepts \cite{manrique2017exploring}. To address this, we apply a concept-weighting strategy that prioritizes contextually and semantically relevant concepts while minimizing noise. Building upon the strategy by Manrique et al. \cite{Manrique2018knowledge}, we propose a transformer-based weighting approach using SBERT \cite{reimers2019sentence} for embedding generation. Our approach ($w_{SBERT}$) assigns weights based on cosine similarity between embeddings of learning material content and Wikipedia article text of the concept, retrieved via the Wikipedia API. The same method is used for related concepts, while Wikipedia categories lacking descriptive text are weighted based on similarity between the learning material content embedding and the category name embedding. This ensures only the most contextually relevant concepts, related concepts, and categories are included in the EduKG.
\section{EduKG Construction Pipelines} \label{oldpipe}
We propose and experiment with two pipelines for EduKG construction in MOOCs, namely Top-down and Bottom-up, discussed below. 
\subsection{Top-down EduKG Construction}
The Top-down approach (Figure \ref{comparison}a) starts by extracting text from the entire PDF learning material, which is passed to the keyphrase extraction module. The keyphrase extractor identifies n keyphrases from the learning material, where n=15*the number of slides in the material, as this formula proved to cover all the possible keyphrases based on experiment. These keyphrases are annotated with DBpedia Spotlight to identify Main Concepts (MCs), which are then weighted by computing the cosine similarity between the MC's Wikipedia article embedding and the learning material's text embedding. Relationships between the MCs and the learning material are stored in a Neo4j database.
This method generates a single EduKG for the entire learning material (LM-EduKG). To provide more granular views, the approach is extended to generate EduKGs for individual slides (Slide-EduKG). Text is extracted from each slide, keyphrases are identified, and the corresponding MCs are checked against the LM-EduKG. If the concept already exists, it is weighted, and relationships are created; otherwise, it is discarded. After covering all slides, concept expansion is applied on the whole EduKG to include related concepts and categories from Wikipedia.
\subsection{Bottom-up EduKG Construction}
The Bottom-up approach (Figure \ref{comparison}b) constructs the EduKG starting from each slide/PDF page of the learning material, with the text extracted from each slide as the initial reference. From each slide, 15 keyphrases are extracted, as more than 95\% of slides contain fewer than 15 keyphrases. These keyphrases are linked to MCs via DBpedia Spotlight, and each concept’s weight is calculated based on the cosine similarity between the SBERT embedding of the concept’s Wikipedia abstract and the SBERT embedding of the learning material text ($w_{LM}$). Additionally, a slide similarity score ($w_{Slide}$) is calculated based on the cosine similarity between the SBERT embedding of the concept’s Wikipedia abstract and the SBERT embedding of the slide text. The final importance of the concept per slide is determined by the sum of the slide similarity score ($w_{Slide}$) and the concept weight ($w_{LM}$).
Relations are established between the MCs and both the slide and the learning material. After annotating each slide, the data is stored in the Neo4j database for immediate access, allowing users to explore the Slide-EduKG even before the entire LM-EduKG is completed. This ensures that users can access partial results while the construction continues. Once all slides are annotated, concept expansion is performed. Lastly, the EduKGs for each slide are aggregated into a comprehensive LM-EduKG. This approach ensures that concepts related to a slide are accurately represented, and any concepts at the slide level are carried over to the learning material level.
\section{Evaluation} \label{eval}
For the evaluation, using our MOOC platform CourseMapper, we conducted an online user study followed by a human annotation study to assess which pipeline produces the most accurate and performant EduKG.
\subsection{Evaluation of EduKG Performance} 
To evaluate the performance of Top-down vs. Bottom-up pipelines, a user study was conducted with 19 participants (11M, 8F) from three different courses taught in our chair. Invitations were sent to 47 individuals, with 19 responding to evaluate 34 learning materials in total.
EduKGs were constructed for different learning materials using both the Top-down and Bottom-up pipelines in CourseMapper. The evaluation involved assessing Precision(P), Mean Reciprocal Rank (MRR), and Mean Average Precision (MAP) for top-k results based on participants' feedback. Participants were introduced to the platform, the research goals, and the evaluation task. They randomly selected a learning material that they were most familiar with, and the corresponding EduKG (consisting of Top-15 MCs) was shown to them. Afterwards, they completed a questionnaire for each Top-down and Bottom-up EduKG.
The questionnaire included questions on: 1) Familiarity with the topic (1: Not familiar, 5: Expert), 2) Relevance of concepts to the material (1 to 15 concepts), 3) Expected concepts not included in the list, and 4) Ranking the concepts from most to least relevant. In addition, users provided feedback on whether the EduKG covered important content, helped them form an understanding of the material, and overall satisfaction with the results.
The results (Table \ref{tab:results}) showed similar performance between both models. However, the bottom-up approach showed slightly better precision at higher k-values and MAP, suggesting that it retrieved more effective information. A T-test revealed no significant differences between the models. In terms of user experience, the bottom-up EduKG was rated more favorably by the participants.
\begin{table}
\caption{Results of the evaluation of Top-down vs. Bottom-up approaches}
\label{tab:results}
\begin{tabular}{|c|c|c|c||c|c|}
\hline
Pipeline  & \multicolumn{3}{c||}{User study}          & \multicolumn{2}{c|}{SRS evaluation}                                                                          \\ \hline
          & P@15           & MRR   & MAP            & Mean Value \begin{math}\mu_s\end{math} & Normal Approximation Value \begin{math}\mu_s \pm \sigma \end{math} \\ \hline
Top-down  & 0.807          & 0.941 & 0.807          & 0.38                                   & 0.38 \begin{math}\pm \end{math} 0.048                              \\
Bottom-up & \textbf{0.812} & 0.941 & \textbf{0.812} & \textbf{0.40}                          & \textbf{0.4 \begin{math}\pm \end{math} 0.049}   \\ \hline                
\end{tabular}
\end{table}
\subsection{Evaluation of EduKG Accuracy}
To assess the accuracy of EduKGs generated using the Top-down and Bottom-up approaches, we employed the Simple Random Sampling (SRS) method by Gao et al. \cite{Gao2019}. This method evaluates the correctness of knowledge graph (KG) triples (subject, predicate, object) through two key tasks: \textit{entity identification} (verifying node meanings using contextual information) and \textit{relationship validation} (ensuring correct links between nodes). 
In this way, accuracy is calculated as the mean of sample judgments. If the margin of error (MoE) exceeds a predefined threshold, additional samples are evaluated until accuracy stabilizes. There are several matrices involved in the calculation of accuracy. The \textit{mean accuracy} (\(\mu_s\)) of a sample set with (\(n_s\)) samples in SRS is computed as:  
\begin{center}
\begin{math}
    \mu_s = \frac{1}{n_s} \sum_{i=1}^{n_s} f(t_i)
\end{math}
\end{center}
where \( f(t_i) \) is 1 for accurate samples and 0 otherwise. The \textit{normal approximation} estimates the accuracy range:
\begin{center}
    \begin{math}
        \mu_s \pm z_{\alpha/2} \sqrt{\frac{\mu_s (1-\mu_s)}{n_s}}
    \end{math}
\end{center}
where \( z_{\alpha/2} \) depends on the confidence interval. The \textit{margin of error (MoE)} quantifies estimation precision and helps to determine the potential amount of error that could occur when using a sample instead of the entire population:
\begin{center}
    \begin{math}
        MoE = z_{\alpha/2} \sqrt{\frac{\sigma^2}{n_s}}
    \end{math}
\end{center}
The evaluation took 4 hours, with two annotators reviewing different random samples. The samples were of type e.g. (Slide, contains, MC), and (LM, contains, MC). The first annotator evaluated 200 samples per model, while the second evaluated 183 Top-down and 180 Bottom-up samples. Evaluation stopped when all criteria were met.
Results (Table \ref{tab:results}) showed that the Bottom-up approach more accurately identified key concepts and their associations with the learning materials and the slides. However, a T-test found no statistically significant difference. Overall, across all evaluations, the Bottom-up pipeline emerged as the most effective and accurate method for EduKG construction in MOOCs.
\section{EduKG Construction with Human-in-the-Loop}
\begin{figure}
\centering
        \includegraphics[width=0.95\textwidth]{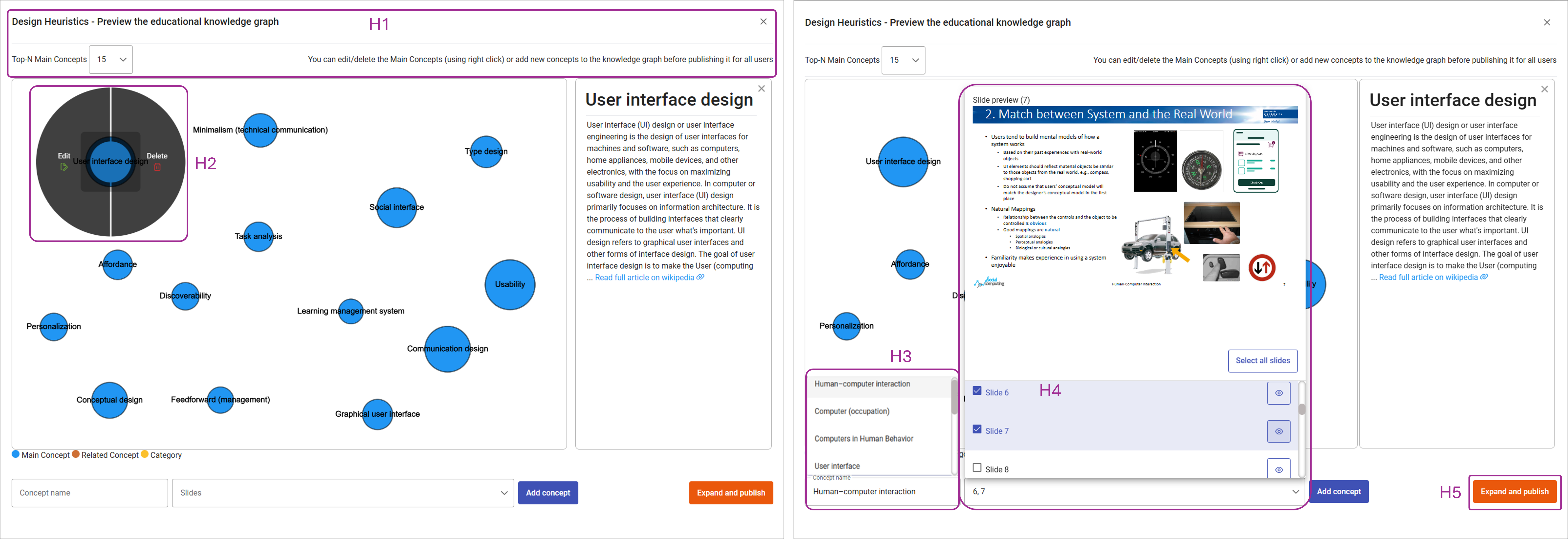}
        \caption{UI of CourseMapper to Preview and Edit the EduKG}
        \label{human}
\end{figure}

Our evaluation revealed that while the Bottom-up approach produced more accurate EduKGs, the overall accuracy was still relatively low. To address this, we integrate a Human-in-the-Loop approach in our pipeline, allowing course creators to review and refine the EduKG before publication. Once the main concepts in the learning material are extracted, instructors can preview them (Figure \ref{human}, H1), edit or remove irrelevant concepts (Figure \ref{human}, H2), and add missing concepts (Figure \ref{human}, H3) and link them to the relevant slide(s) of the learning material (Figure \ref{human}, H4). After finalizing the edits (Figure \ref{human}, H5), the concept expansion step is applied and the verified EduKG is published and presented to learners. This process guarantees accurate EduKGs while striking a balance between automation and human expertise. Moreover, it ensures that learners receive an accurate and instructor-approved EduKG, minimizing the risk of disseminating incorrect or incomplete information.
\section{Conclusion} \label{concl}
In this paper, we explored Top-down vs. Bottom-up approaches for the automatic construction of Educational Knowledge Graphs (EduKGs) in the MOOC platform CourseMapper. We evaluated both approaches and found the Bottom-up approach to be more accurate and effective for EduKG construction at various levels. To further improve EduKG accuracy, we proposed a human-in-the-loop approach, allowing expert refinement while maintaining efficiency. 
\begin{credits}
\subsubsection{\discintname}
The authors have no competing interests to declare that are
relevant to the content of this article. 
\end{credits}
\bibliographystyle{splncs04}  
\bibliography{references}     
\end{document}